\documentclass[11pt,a4paper]{article}

\usepackage[margin=1in]{geometry}
\usepackage{amsmath,amssymb}
\usepackage{graphicx}
\usepackage{booktabs}
\usepackage[authoryear,round]{natbib}
\usepackage{hyperref}
\usepackage[nameinlink]{cleveref}
\usepackage{pdfpages}

\newcommand{\Description}[1]{}

\graphicspath{{images/}}

\title{Algorithmic Fairness Perceptions in the Global South: Evidence from Bangladesh on Ride-Sharing, Beauty Filters, and Large Language Models}

\author{%
  Ahmed Abdal Shafi Rasel\thanks{Corresponding author: \texttt{ahmed.shafi@ewubd.edu}} \\
  Ahmed Mustafa Amlan \\
  Tasmim Shajahan Mim \\[6pt]
  \normalsize Department of Computer Science and Engineering, East West University, Dhaka, Bangladesh
}
\date{}

\begin{document}

\maketitle

\begin{abstract}
Algorithmic fairness research comes almost entirely out of North America and Western Europe, so we know surprisingly little about how people elsewhere judge the algorithms they already rely on every day. We asked people in Bangladesh directly: a bilingual (Bangla and English) survey of 199 participants rated fairness across three everyday scenarios --- ride-sharing prices that shift with context, AI beauty filters that reshape appearance, and large language models that handle cultural values differently than a human would.

Four patterns stood out. Context changes the verdict even when the outcome doesn't: a 20\% price surge during a medical emergency feels less fair than the identical surge on a casual trip (2.00 vs.\ 2.17 on a 5-point scale, Wilcoxon $p=.006$) --- a small effect, close to what this sample can reliably detect, and uneven across income groups (largest among middle-income participants, though we did not test that interaction formally); people already view surge pricing critically in general, and context sharpens the judgment rather than creating it. Demand for transparency, consent, and real user control is close to universal: 85.7\% to 90.3\% of participants want these protections regardless of gender, income, or prior awareness of algorithmic bias, so transparency reads as a default-feature question, not a question of who deserves it. Beauty-filter harm tracks with self-image more than with social pressure people can easily name; feeling personally affected predicts reduced self-confidence tightly ($R^2=.30$, $p<.001$), though the underlying scale's direction is inferred from context rather than guaranteed by labeled anchors. And among participants who noticed specific instances of LLM cultural bias, a meaningful share could not say why when asked to elaborate --- recognizing bias and being able to articulate it turn out to be different skills.

Outcome-only fairness metrics would miss every one of these patterns.
\end{abstract}

\noindent\textbf{Keywords:} Algorithmic fairness, Algorithmic bias, Global South, Bangladesh, Ride-sharing, Beauty filters, Large language models, Body image, Algorithmic transparency, Digital rights

\section{Introduction}
\label{sec:introduction}

\subsection{Why Study Fairness Perceptions in Bangladesh?}

Algorithms now decide who gets a loan, what content fills your feed, what you pay for a ride, and which opportunities show up at all. The upside is real: speed, personalization, scale. So is the downside, and a substantial literature has grown up around measuring and reducing bias in automated decisions.

Almost all of that literature comes from North America and Western Europe. We know surprisingly little about how people elsewhere actually judge algorithmic fairness --- and that gap has real consequences: when systems are designed without understanding local context, they misalign with the people using them, and the harm lands hardest on those already most exposed to it.

The gap is especially stark for Bangladesh, where smartphone and internet access is expanding fast: cheap data and ride-sharing apps, video platforms, photo filters, and targeted advertising have moved into daily life well ahead of any empirical research on how Bangladeshi users actually perceive fairness in the systems they now use constantly.

\subsection{What We Asked}

This study asks one central question: how do Bangladeshi users judge fairness in the algorithmic decisions they run into every day? To answer it, we ran a bilingual (Bangla and English) survey built around realistic scenarios from three domains: ride-sharing apps that charge different users different prices and raise prices during emergencies, AI beauty filters that alter appearance toward narrow standards, and large language models that handle cultural values differently than a human would.

Four research questions guide the work: Can users recognize when an algorithm might be unfair or biased? How do perceived harms and benefits shape fairness judgments? How do those judgments shift across domains and situations? And what transparency, control, and consent mechanisms do users actually expect?

\subsection{Why This Matters}

This matters for three practical reasons. It is a design problem: systems built without local input tend to serve their users badly, since fairness cannot be assessed in the abstract, only checked against how the people affected actually think and feel. It is a policy problem: Bangladesh and comparable countries are writing AI regulations right now, and those should rest on evidence about what users actually want, not assumptions imported from elsewhere. And it is a literature gap: fairness scholarship still leans heavily on North America and Western Europe, and different cultural and economic contexts surface considerations current frameworks simply miss.

\subsection{What We Found in Brief}

Four patterns run through the results in \Cref{sec:results} ($N=199$). Fairness judgments track situational stakes more than the bare outcome: an identical price increase is judged more harshly in a medical emergency than in a routine trip, though the effect is modest. Demand for transparency, consent, and user control is broad and largely unmoored from demographic background. Beauty-filter harm shows up more in how people see themselves than in social pressure they can name. And among participants who report noticing specific instances of LLM cultural bias, many cannot say why when asked to elaborate.

\subsection{Our Contributions}

This work makes three contributions. The first is empirical: evidence on fairness perceptions from Bangladesh that extends the literature beyond Western contexts. The second is conceptual: evidence that context shapes fairness judgments in ways outcome-focused metrics miss. The third is methodological: a fully reproducible analysis pipeline, complete with de-identified data, codebook, and a single re-runnable script, that other researchers can adapt for similar work in under-resourced settings.

The rest of the paper lays these findings out and works through what they mean for system design, policy, and future research.

\section{Background and Related Work}
\label{sec:literature-review}

\subsection{How People Judge Fairness in Algorithms}

Fairness is not only about whether outcomes come out evenly distributed. \citet{yurrita2023} show people judge it through several channels at once, weighing whether they feel adequately informed, whether the process seems reasonable, and whether the overall outcome feels right. Personal stakes shape these judgments too: people tend to view outcomes that favor them as fairer, even when the algorithm is described as biased \citep{wang2020factors}.

That challenges purely technical approaches to fairness. \citet{hosseini2024} argue that mathematical criteria alone cannot capture real human behavior. \citet{binns2018} show users trust systems more when the process feels fair, not just the outcome; in hiring scenarios, even small improvements in perceived fair treatment produced large gains in trust \citep{zhou2021understanding}.

How an outcome gets explained matters too, though maybe not the way you would expect: \citet{shulner2022} find users care more about whether an outcome feels fair than about how thoroughly it is explained, and a thorough explanation bolted onto an unfavorable outcome does almost nothing for perceptions. Platform design itself can reshape fairness norms --- \citet{chong2024} show that technology-mediated marketplaces make it harder for consumers to act on fairness norms that existed before the platform arrived.

\subsection{Fairness Metrics and Their Limits}

If perception is this multidimensional, the technical metrics built to formalize fairness face an uphill fight. Different fairness criteria often conflict outright: \citet{buyl2024} show fairness and accuracy can trade off against each other, and in hiring specifically, \citet{poe2024} demonstrate that satisfying a statistical fairness criterion does not guarantee compliance with non-discrimination law --- a system can pass every technical fairness test and still violate legal or ethical standards.

\citet{corbett2023} make a sharper point: enforcing demographic parity without regard to risk distributions can worsen inequality, so fairness interventions have to match the domain's specific risk profile. \citet{starke2022} add that different demographic groups notice different kinds of errors, which means fairness assessments should incorporate social context beyond group membership alone.

\subsection{Who Judges Fairness and How}

User background matters for how these judgments get made. Age, gender, education, and prior experience all shape how people read algorithmic decisions \citep{wang2020factors,grgic2018,kleanthous2022}, and technical literacy plays a role too: \citet{yurrita2023} find participants who rated their AI knowledge higher also perceived systems as more honest.

But domain knowledge matters too, sometimes more than technical skill. \citet{rozado2023} show that detecting political bias in large language models takes political knowledge, not just technical skill, and \citet{nielsen2024} find even environmentally-aware participants systematically underestimate carbon-footprint inequality --- awareness alone does not guarantee accurate perception of systemic patterns.

Fairness judgments are also relational, not just individual. Interviewing content creators about platform moderation, \citet{ma2022} find fairness gets assessed against similar peers, against the platform's past decisions, and against whether users had any voice in the outcome, not just against whether outcomes were statistically equal.

\subsection{Large Language Models and Bias}

Generative AI introduces a version of this problem that predictive systems never had. \citet{rozado2023} gave ChatGPT standardized political-orientation tests and found consistent left-leaning bias across most of them; unlike bias in predictive systems, bias in generative text shapes what gets written, which views get represented, and whether competing perspectives receive equal treatment.

Political context also shapes how people read that bias: cross-ideological audiences may label the exact same model output as biased or as appropriately balanced, depending on their own priors \citep{haider2024}, which complicates technical bias assessment since the same output can be judged differently by different groups.

\subsection{Beauty Filters and Appearance Bias}

AI-mediated beauty tools reinforce narrow beauty standards directly. \citet{riccio2023} audit augmented-reality beauty filters and document that they systematically embed Eurocentric norms, lightening skin tone and reshaping features toward a single ideal --- technical evidence for the bias survey respondents in this study report perceiving.

\citet{rahman2025} analyze beauty-enhancement discourse across social media and find a dual effect: these tools support self-esteem for some users while reinforcing harmful standards for others, which lines up with this study's own finding that beauty-filter harm operates through internalized effects on self-perception.

\subsection{What's Missing}

Most fairness-perception research comes out of Western contexts, but a smaller Global South literature is directly relevant here. \citet{ramesh2022} study how platform-user power relations shape algorithmic accountability among financially stressed instant-loan borrowers in India, finding that structural power imbalances, not just information gaps, determine whether users can meaningfully contest algorithmic decisions. \citet{santos2020} study dynamic pricing and price-fairness perceptions among Uber riders directly, giving this paper's own flagship finding a closer predecessor than the general algorithmic-fairness literature above supplies.

Two things distinguish this study within that smaller literature: the context, and the method. \citet{ramesh2022} and \citet{santos2020} are themselves exceptions to a Western-dominated field; this study adds Bangladesh to a list that is still short. On method, much of the domain-specific work cited earlier in this section (\citet{riccio2023}'s beauty-filter audit, \citet{rahman2025}'s discourse analysis, \citet{ma2022}'s interviews with content creators) measures bias technically or reads it out of existing discourse rather than asking people directly, at scale, how they judge it. This study instead combines a 199-person quantitative survey with qualitative coding of open-ended responses, capturing both how widely a judgment is held and, for a subset of respondents, why. \Cref{sec:discussion} returns to what specifically, if anything, looks different from the work cited here.

\section{Methods}
\label{sec:methodology}

\subsection{Study Design}

We used a bilingual survey with realistic scenarios. Every participant saw three scenarios describing algorithmic systems they might run into in daily life: ride-sharing apps with variable pricing, beauty filters that alter appearance, and AI systems that handle cultural values. Each scenario carried questions about awareness, fairness judgments, perceived harms, and what improvements participants wanted.

The survey appeared side by side in Bangla and English. Native speakers reviewed the Bangla translation. Participants could respond on paper or digitally, with research assistants available to help them work through the material.

\subsection{Participants}

We collected 201 responses between 14 July 2025 and 1 February 2026. Two declined consent, leaving 199 in the analytic sample. Recruitment went through university networks, professional associations, and community organizations.

\begin{table}
\centering
\caption{Who Participated ($N=199$)}
\label{tab:demographics}
\begin{tabular}{llrr}
\toprule
\textbf{Characteristic} & \textbf{Category} & \textbf{n} & \textbf{\%} \\
\midrule
Age & Under 18        & 14 & 7.0\%  \\
    & 18--20          & 36 & 18.1\% \\
    & 21--23          & 50 & 25.1\% \\
    & 24--26          & 73 & 36.7\% \\
    & 27--30          & 14 & 7.0\%  \\
    & Above 30        & 12 & 6.0\%  \\
\midrule
Gender & Male          & 131 & 65.8\% \\
       & Female        & 68  & 34.2\% \\
\midrule
Residence & Major City      & 128 & 64.3\% \\
          & Town/Small City & 49  & 24.6\% \\
          & Rural/Village   & 22  & 11.1\% \\
\midrule
Income & Lower-income  & 46  & 23.1\% \\
       & Middle-income & 130 & 65.3\% \\
       & Upper-income  & 23  & 11.6\% \\
\midrule
Education & SSC or below  & 23 & 11.6\% \\
          & HSC           & 66 & 33.2\% \\
          & Bachelor's    & 99 & 49.7\% \\
          & Master's      & 11 & 5.5\%  \\
\midrule
Technology familiarity & Very Low / Low  & 24 & 12.1\% \\
                        & Moderate       & 93 & 46.7\% \\
                        & High / Very High & 82 & 41.2\% \\
\midrule
Prior awareness of algorithmic bias & Yes & 132 & 66.3\% \\
                                   & No  & 67  & 33.7\% \\
\bottomrule
\end{tabular}
\end{table}

The sample skews young, urban, male, and middle-income. That is typical for convenience samples in digital-literacy research. It gives us real insight into how educated, digitally-engaged Bangladeshis perceive algorithmic systems, but it limits generalization to the broader population. We flag this honestly.

\subsection{Supplementary Follow-Up Sample}
\label{subsec:followup-sample}

A separate, smaller sample ($N=47$) came through a different survey focused only on ride-sharing, recruited primarily from ride-sharing drivers. Its collection window (24 Aug -- 10 Dec 2025) sits entirely inside the primary survey's own window (14 Jul 2025 -- 1 Feb 2026) rather than starting after it closed, so the two samples were fielded concurrently; we cannot verify their independence (a driver could in principle have also answered the primary survey). This follow-up used a different consent model that collected names and, optionally, email addresses and signatures; those fields are dropped before analysis and never appear in any output. This sample is supplementary and is not pooled with the main $N=199$ in any primary analysis; \Cref{sec:results} reports a naive pooled sensitivity check available in the released pipeline for transparency, not as a headline statistic.

\subsection{Analysis Approach}

All analyses were run in Python using pandas, scipy.stats, and statsmodels. The complete analysis script, the codebook mapping survey columns to variable names, and the results are all available with this paper (see the Data and Code Availability section).

\textbf{Handling missing data.} Each statistic uses all participants who answered the specific items involved (pairwise deletion). Effective N varies slightly by question.

\textbf{Statistical tests.} We checked normality with Shapiro-Wilk tests. All Likert-scale measures deviated significantly from normal ($p<10^{-7}$), so Wilcoxon signed-rank tests serve as the primary result for paired comparisons. Paired t-tests appear alongside for comparability. For group comparisons, we used chi-square tests for categorical data and ANOVA (with Levene's test for equal variances) for continuous measures.

\textbf{Effect sizes.} Cohen's d accompanies paired comparisons. Phi coefficients are computed for every chi-square test and are available in the released results (Data and Code Availability section); we report them in-text only where they carry the argument, since the manuscript quotes chi-square and p-values for the RQ4 tests directly rather than every derived statistic. Minimum detectable effect size at 80\% power is reported for the manuscript's primary, pre-specified comparison (the ride-sharing context effect); it is also computed for every other paired and chi-square comparison and is available in the released results, so a reader who wants to weigh a specific null result against its detection floor can do so even where we do not quote the number in-text. We report 95\% confidence intervals (t-distribution for paired mean differences, Fisher $z$ for Spearman and Pearson correlations) alongside the manuscript's primary paired comparisons and correlations, so a reader can see the plausible range around each point estimate rather than only its significance.

\textbf{Multiple comparisons.} Bonferroni correction is applied within families of related tests (four RQ4 tests each by income, gender, and prior algorithmic-bias awareness; three technical-familiarity correlation tests), not across the entire analysis. Three further RQ4 families (by education, occupation, and residence; \Cref{sec:results}) are exploratory, post hoc additions rather than part of this pre-specified set, each corrected within its own family of four for the same reason.

\textbf{Data quality.} The released raw export (\texttt{survey.csv}) has 23 unlabeled columns (57--79 in the raw file) with no header text. Most of this block's content matches an unrelated survey about educational-content recommendation, left over from shared form infrastructure and not part of this study's instrument (\Cref{app:survey-instrument}); however, one column inside that same unlabeled block (column 59) carries this study's own beauty-filter ``why'' follow-up (\Cref{subsec:qualitative-results}, Item C), so the block is not uniformly foreign data and we did not exclude it wholesale -- each of the 23 columns was checked individually for content before being included in or excluded from analysis, rather than dropped as a range. One open-text response to the LLM cultural-exclusion item was identified as a pasted large-language-model output rather than a personal response and was excluded from that item's counts in \Cref{subsec:qualitative-results} and \Cref{sec:results}; separately, two non-substantive responses to a different LLM open-text item (``don't know'' and ``test'') were excluded from that item's coded themes, but bare ``no''-type answers to the cultural-exclusion item itself are counted as complete answers, not excluded (\Cref{sec:results}). No screening beyond this was applied to the closed-form items.

\section{Results}
\label{sec:results}

All statistics come from the analysis pipeline described in \Cref{sec:methodology}. Each result states its N, reflecting pairwise-available responses for that specific question.

\subsection{Finding 1: Context Changes How People Judge Fairness}

\subsubsection{Most people know prices vary}

Most participants (69.3\%, $n=199$) already knew ride-sharing apps can show different prices to different users. Awareness of differential pricing was higher in the supplementary driver sample (91.5\%) --- but that sample is not representative of general users.

\subsubsection{Emergency pricing feels less fair than casual pricing}

We asked participants to rate the fairness of a 20\% price increase in two situations: during a medical emergency, and during a casual trip. The same increase got judged differently.

\begin{figure}[h]
    \centering
    \includegraphics[width=0.75\linewidth]{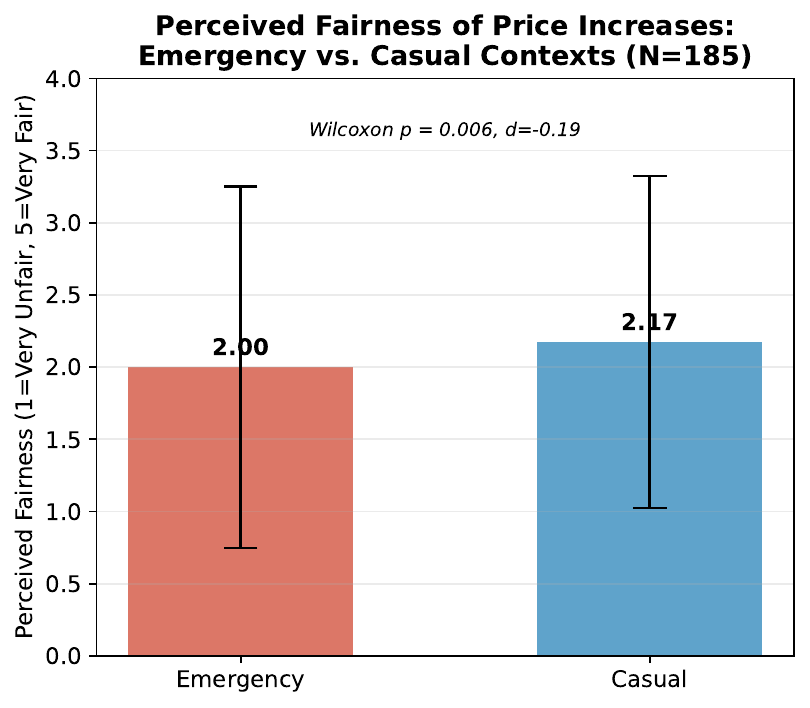}
    \caption{Fairness ratings for the same 20\% price increase in emergency vs.\ casual contexts ($N=185$). Lower scores mean less fair.}
    \Description{Bar chart comparing mean fairness ratings for a 20 percent price increase in an emergency context (mean 2.00) versus a casual context (mean 2.17), both on a 1-to-5 scale where lower is less fair; the emergency-context bar is shorter.}
    \label{fig:rideshare_context}
\end{figure}

The emergency-context increase ($M=2.00$) was rated less fair than the casual-context increase ($M=2.17$; mean difference 0.17, 95\% CI $[0.04, 0.30]$). The difference is statistically significant (Wilcoxon $p=.006$, paired $t=-2.64$, $p=.009$); the effect size is small (Cohen's $d=-0.19$) and sits just below this sample's own minimum reliably-detectable effect for this comparison at 80\% power ($d=0.207$, $n=185$) --- a real caveat, not a rounding note, since it means an effect of exactly this magnitude is close to what this sample could fail to detect even if the true population effect were somewhat larger or smaller. The confidence interval is consistent with that caveat: it excludes zero, but only just, and its lower bound (0.04) is a very small effect in its own right.

This comparison also carries a confound we did not control for: the emergency item was always presented immediately before the casual item (\Cref{app:survey-instrument}, Scenario 1, items 9--10), with randomization neither implemented nor verifiable from the export. An order or contrast effect --- respondents anchoring the second rating against the first --- is a plausible alternative or additional explanation for part of this gap, one this design cannot separate from the emergency-versus-casual manipulation itself.

Both ratings sit well below the midpoint of the scale. People view surge pricing critically in general --- context sharpens that judgment, it does not create it. Perceived harm from the pricing scenario overall also correlates strongly with reduced trust in the platform (Spearman $\rho=.599$, 95\% CI $[.50, .68]$, $p<.001$, $n=199$): participants who feel more harmed by differential pricing report a correspondingly larger hit to their trust in the platform, not an independent reaction.

\subsubsection{Income level shapes the context effect}

Lower-income participants rated both contexts as similarly unfair, with little gap between emergency ($M=2.05$) and casual ($M=2.12$). Middle-income participants showed the largest gap between contexts ($M=2.01$ vs.\ $M=2.24$). Upper-income participants rated both contexts most favorably ($M=1.87$ vs.\ $M=1.91$).

These are descriptive subgroup means. The upper-income subgroup ($n=23$) is too small to detect anything but large effects. Treat the patterns as suggestive rather than conclusive.

\begin{figure}[h]
    \centering
    \includegraphics[width=0.75\linewidth]{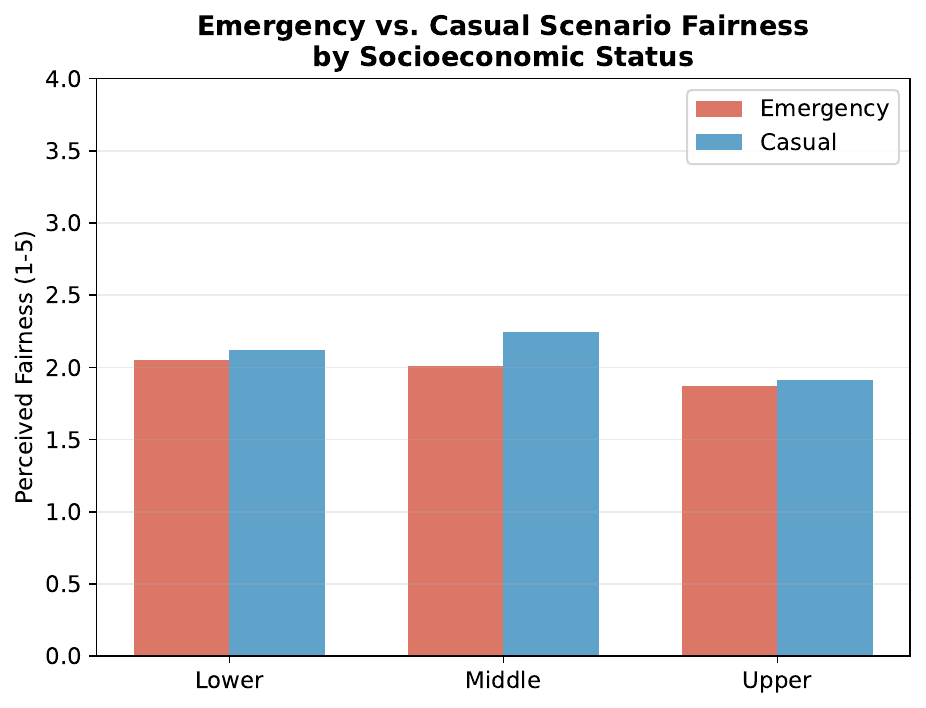}
    \caption{Emergency vs.\ casual fairness ratings by income group.}
    \Description{Grouped bar chart of emergency versus casual fairness ratings across three income groups: lower-income (2.05 vs.\ 2.12), middle-income (2.01 vs.\ 2.24, the largest gap), and upper-income (1.87 vs.\ 1.91, the smallest gap and most favorable ratings overall).}
    \label{fig:socioeconomic_moderation}
\end{figure}

\subsubsection{A driver sample showed different patterns}
\label{subsec:followup-results}

A separate sample of 47 ride-sharing drivers showed no significant gap between emergency and casual pricing ($M=2.57$ vs.\ $M=2.60$, $p=.64$); the sample is small and skewed (100\% male, older, predominantly rural), and not representative of Bangladeshi users broadly. As a transparency check rather than a primary result, naively pooling this sample with the main $N=199$ ($n=232$ combined) still shows a significant, same-direction context effect (Wilcoxon $p=.007$, Cohen's $d=-0.159$) --- the pooled sensitivity check available in the released pipeline does not flip the finding's direction, though we do not treat pooling as a valid design given the two samples' different instruments and consent models.

Drivers rated the pricing scheme fair overall more often than passengers did (61.7\% vs.\ 26.1\%, $p<.001$). We read this cautiously rather than as a clean group comparison: the driver and passenger samples used different consent models, different recruitment channels, and overlapping (not sequential) fielding windows (\Cref{subsec:followup-sample}), and the driver instrument itself asked drivers to answer the ride-sharing items from the passenger's perspective (``imagine you urgently need a ride to the hospital,'' \Cref{app:followup-instrument}), not as drivers pricing a fare. Any explanation resting on drivers' and passengers' different economic positions --- drivers earn from fares, passengers pay them --- is therefore a plausible reading of the raw percentages, not a mechanism this survey design can confirm or attribute to that specific cause over instrument and sampling differences.

\subsection{Finding 2: Beauty Filters Harm Self-Image More Than Social Pressure}

\subsubsection{Most people recognize beauty-filter bias}

Most participants (73.2\%, $n=198$) already knew beauty filters promote narrow beauty standards.

\subsubsection{Self-image impact exceeds social pressure}

We asked participants to separate two effects: how filters affect their self-image and confidence, versus how much social pressure they feel to use filters.

Self-image impact ($M=2.97$) was significantly higher than social pressure ($M=2.46$; mean difference 0.51, 95\% CI $[0.28, 0.73]$; Wilcoxon $p<.001$, Cohen's $d=0.32$).

This comparison rests on two items that differ in more than construct: the self-image item asks about filters' effect in general (``How do such filters affect confidence or self-image?''), while the social-pressure item is explicitly first-person (``Do \emph{you} feel social pressure to use beauty filters to fit in?''; \Cref{app:survey-instrument}). That wording asymmetry is a plausible alternative explanation for the gap on its own: a well-documented pattern in media-effects research is that people rate a medium's effect on people in general higher than its effect on themselves specifically (a third-person effect), independent of any difference between self-image harm and social pressure as constructs. We cannot rule this out with the items as asked, and flag it as a live alternative explanation rather than a settled harm mechanism.

Harm here operates mainly through internalized effects on how people see themselves, on the reading above, rather than through obvious external pressure.

\begin{figure}[h]
    \centering
    \includegraphics[width=0.75\linewidth]{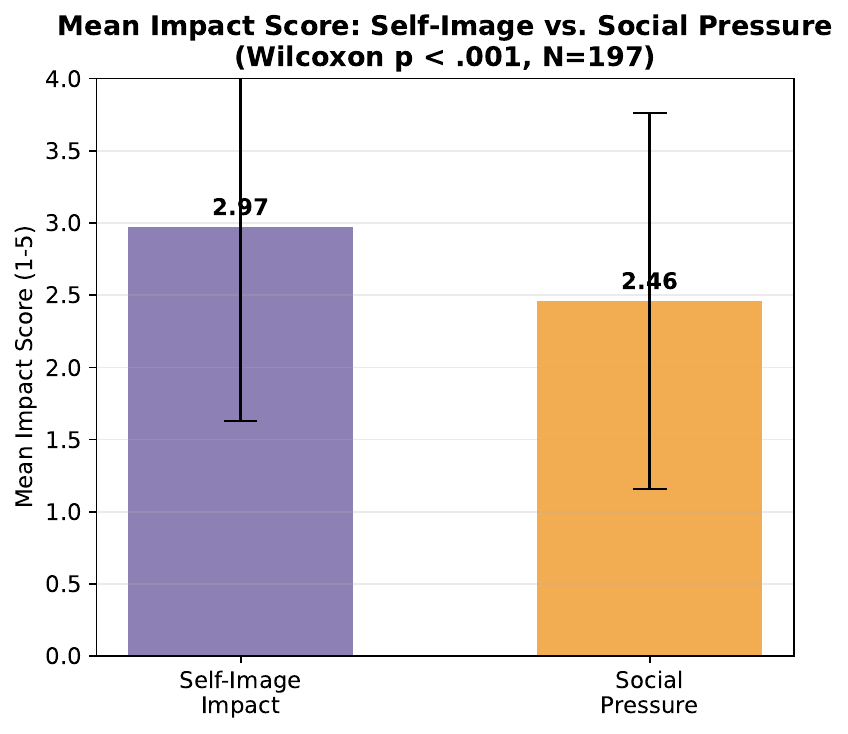}
    \caption{Self-image impact vs.\ social pressure from beauty filters.}
    \Description{Bar chart comparing mean self-image impact (2.97) to mean social pressure (2.46) from beauty filters on a 1-to-5 scale; the self-image bar is taller.}
    \label{fig:beauty_selfimage_socialpressure}
\end{figure}

Feeling personally affected by filters tracks closely with reduced self-image. Higher personal affect predicts stronger self-image impact ($R^2=.30$, $r=.545$, 95\% CI $[.44, .64]$, $p<.001$). The relationship does not differ by gender (interaction $p=.945$).

\textbf{Important caveat.} This survey did not measure actual filter-use frequency. What we show is a relationship between two self-reported impact dimensions, not a direct exposure-response mechanism. Keep that limit in mind when reading the result.

\subsubsection{Qualitative responses describe similar harms}

Open-ended responses to this scenario described struggling to accept one's unfiltered appearance, and in one case explicitly racialized experiences. Participants named authenticity loss rather than social comparison as the mechanism. (See \Cref{subsec:qualitative-results} for coded examples.)

\subsection{Finding 3: Demand for Transparency and Control Is Widespread}

\subsubsection{Most participants want user-control mechanisms}

85.7\% to 90.3\% of participants supported each user-control mechanism. Specifically, 90.3\% want platforms to explain how their systems work ($n=196$), 85.7\% want systems to ask for preferences before making decisions ($n=196$), 86.7\% want to customize how systems work for them ($n=196$), and 88.0\% want clear warnings when outputs may be biased ($n=192$).

\subsubsection{Support does not differ by demographic group, gender, or prior awareness}

Demand for these mechanisms showed no significant relationship with income, gender, or prior awareness of algorithmic bias, across all four measures. By income: customization $\chi^2(2)=0.19$, $p=.91$; explanation $\chi^2(2)=1.56$, $p=.46$; prior consent $\chi^2(2)=0.07$, $p=.97$; bias indicators $\chi^2(2)=1.83$, $p=.40$ (all four of these by-income tests have a minimum expected cell count below 5 given the small upper-income subgroup, so the chi-square approximation is less reliable here than for the by-gender and by-awareness tests below; we report them for completeness but weight them less). By gender: customization $\chi^2(1)=1.35$, $p=.25$; explanation $\chi^2(1)=0.00$, $p=1.0$; prior consent $\chi^2(1)=0.00$, $p=.98$; bias indicators $\chi^2(1)=0.02$, $p=.89$.

\begin{figure}[h]
    \centering
    \includegraphics[width=0.75\linewidth]{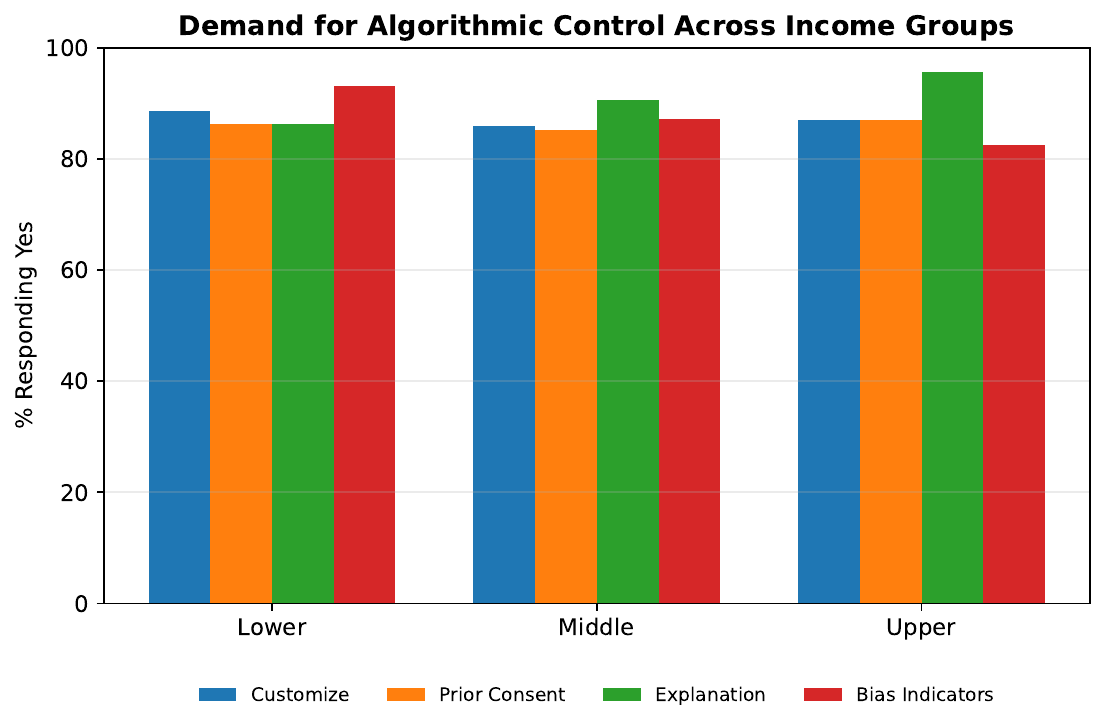}
    \caption{Support for user-control mechanisms by income level.}
    \Description{Grouped bar chart showing support (as a percentage) for four user-control mechanisms -- explanation, customization, prior consent, bias indicators -- broken down by lower-, middle-, and upper-income groups, ranging from 82.6 percent (bias indicators, upper-income) to 95.7 percent (explanation, upper-income), with no consistent income gradient across the four mechanisms.}
    \label{fig:rq4_by_ses}
\end{figure}

What predicts support, then? Not prior awareness either, on this data: comparing participants who already knew about algorithmic bias against those who did not, none of the four by-awareness comparisons reach significance (customization $\chi^2(1)=0.71$, $p=.40$; explanation $\chi^2(1)=0.38$, $p=.54$; prior consent $\chi^2(1)=0.92$, $p=.34$; bias indicators $\chi^2(1)=1.78$, $p=.18$; all remain non-significant after Bonferroni correction within this family of four). Aware participants do show a consistently higher raw support rate on every measure (for example 88.5\% vs.\ 83.1\% on customization, 90.6\% vs.\ 82.8\% on bias indicators), so the direction is not random, but the difference does not clear statistical significance in this sample. Read together with the income and gender results above, the more accurate summary is that we did not find a predictor of RQ4 support among the demographic and awareness variables we measured --- support looks close to universal rather than concentrated in any subgroup we could identify.

\begin{figure}[h]
    \centering
    \includegraphics[width=0.75\linewidth]{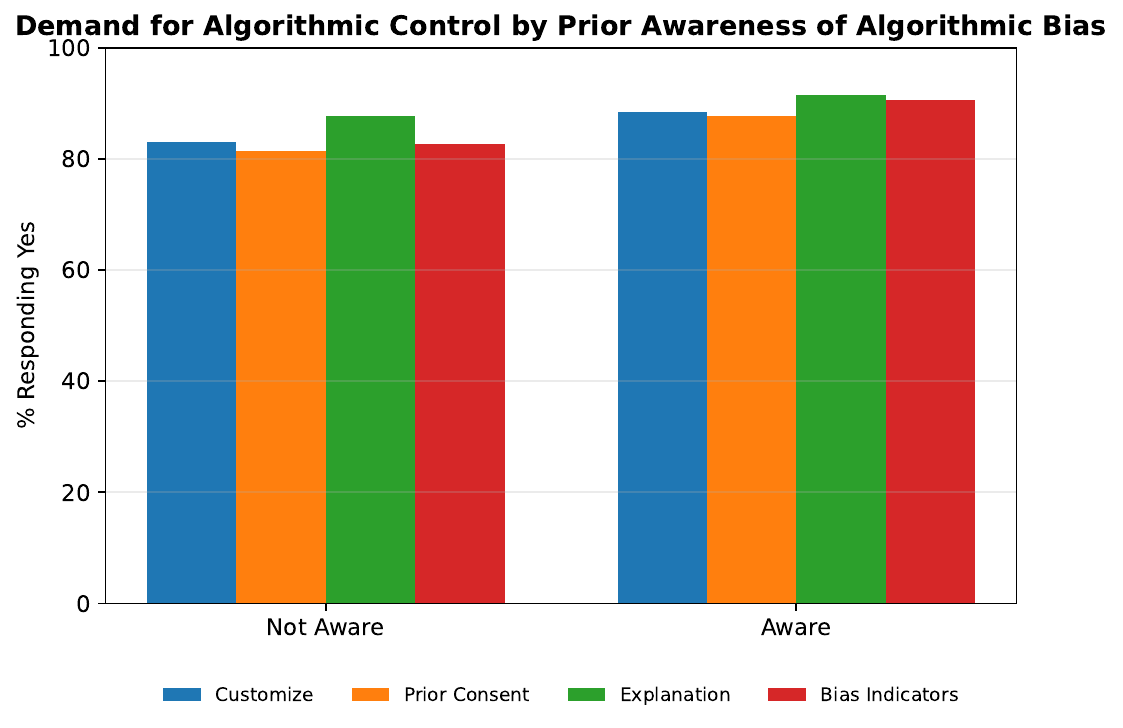}
    \caption{Support for user-control mechanisms by prior awareness of algorithmic bias.}
    \Description{Grouped bar chart showing support (as a percentage) for the same four user-control mechanisms broken down by whether respondents were previously aware of algorithmic bias. Aware respondents show a consistently higher bar on all four mechanisms than not-aware respondents, but none of the four gaps reach statistical significance.}
    \label{fig:rq4_by_awareness}
\end{figure}

We hold that summary alongside a competing explanation the instrument cannot rule out. All four RQ4 items are unipolar, low-cost, socially desirable propositions (``Should platforms explain how their system works?''), with no reverse-worded item and no attention check anywhere in the instrument (\Cref{sec:methodology}). At 85.7--90.3\% agreement there is very little variance left for income, gender, or awareness to explain, so the twelve null subgroup tests above are close to guaranteed by the response distribution itself and cannot, on their own, distinguish a genuinely universal preference from acquiescence bias toward agreeable-sounding statements. We report the finding as we measured it, but a future instrument with a reverse-worded or cost-trading item (e.g.\ transparency traded against convenience or latency) would let a reader tell these apart.

\begin{figure}[h]
    \centering
    \includegraphics[width=0.75\linewidth]{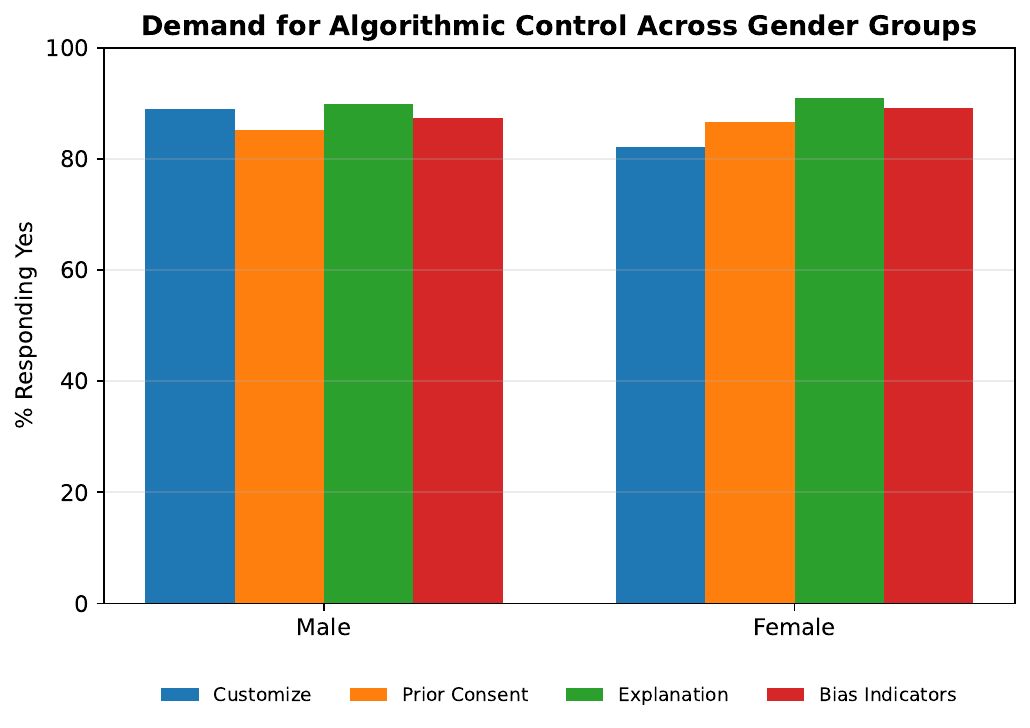}
    \caption{Support for user-control mechanisms by gender.}
    \Description{Grouped bar chart showing support (as a percentage) for the same four user-control mechanisms broken down by male and female respondents, with male and female bars closely matched for every mechanism and no significant gender gap.}
    \label{fig:rq4_by_gender}
\end{figure}

\subsubsection{Technical familiarity and perceived harm}

Self-reported technology familiarity correlates with perceived harm inconsistently across the three domains, not as a single general pattern. In ride-sharing, higher technical familiarity predicts \emph{higher} reported harm from differential pricing (Spearman $\rho=.217$, 95\% CI $[.08, .35]$, $p=.002$, $n=199$), the one relationship of the three that survives Bonferroni correction within this family of three tests (corrected $p=.006$); the confidence interval is wide and its lower bound is a weak correlation, so treat the size of this effect as uncertain even where the direction is not. In beauty filters and LLM cultural bias, the same relationship is not significant (beauty filters: $\rho=-.028$, $p=.69$, $n=197$; LLM: $\rho=.116$, $p=.11$, $n=197$; both non-significant after correction).

\begin{figure}[h]
    \centering
    \includegraphics[width=0.75\linewidth]{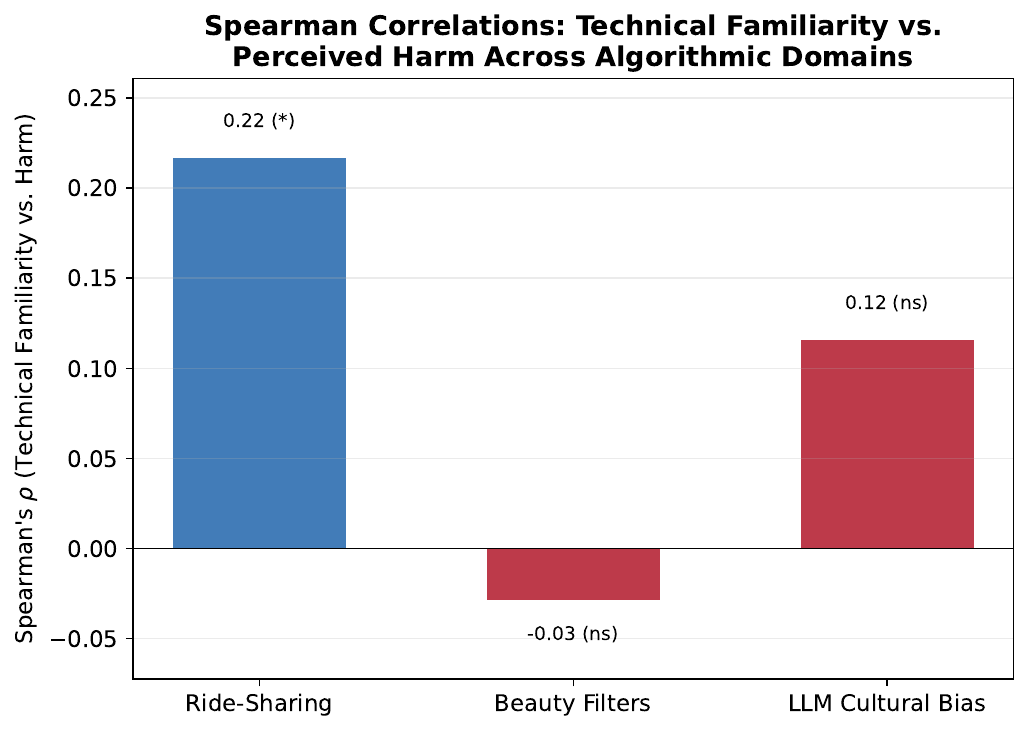}
    \caption{Spearman correlations between technical familiarity and perceived harm, by domain.}
    \Description{Bar chart of Spearman's rho for technical familiarity versus perceived harm across three domains: ride-sharing (rho=0.22, significant), beauty filters (rho=-0.03, not significant), and LLM cultural bias (rho=0.12, not significant).}
    \label{fig:tech_familiarity_harm}
\end{figure}

This is the one place in our data where more technical knowledge tracks with a stronger reaction rather than a flatter one, and it is domain-specific: more technically familiar participants report \emph{more} harm from ride-sharing price discrimination specifically, not more harm in general. We do not have a mechanism for this from the survey alone --- one plausible reading is that technically familiar users better understand how personalized pricing is computed and find that understanding itself unsettling, but this survey cannot distinguish that from other explanations.

\subsubsection{Exploratory Subgroup Moderators: Education, Occupation, and Residence}

Beyond the pre-specified income, gender, and awareness comparisons above, the dataset also supports RQ4 support tests by education level, occupation, and urban/rural residence. We report these as exploratory, post hoc additions, not part of the paper's pre-specified test families, and all three carry the same small-subgroup caveat as the income comparisons: expected chi-square cell counts fall below the conventional adequacy threshold throughout (education's smallest group, Master's, is $n=11$; occupation's smallest, an undocumented ``Faculty'' write-in category, is $n=4$; residence's rural/village group is $n=22$).

\begin{table}[h]
\centering
\caption{RQ4 Support by Education, Occupation, and Residence (Exploratory, Uncorrected $p$ / Bonferroni-Corrected $p$)}
\label{tab:rq4-exploratory-moderators}
\small
\begin{tabular}{lrrrr}
\toprule
\textbf{Moderator} & \textbf{Customize} & \textbf{Explain} & \textbf{Prior Consent} & \textbf{Bias Ind.} \\
\midrule
Education (4 groups) & .120 / .482 & .444 / 1.00 & .186 / .743 & .347 / 1.00 \\
Occupation (5 groups) & .603 / 1.00 & .671 / 1.00 & .800 / 1.00 & .556 / 1.00 \\
Residence (3 groups) & .040 / .159 & .726 / 1.00 & .028 / .112 & .124 / .497 \\
\bottomrule
\end{tabular}
\end{table}

Education and occupation show no differences at either the uncorrected or corrected level. Residence is the one place a raw $p$-value crosses .05 (customize, prior consent) before correction, with lower support in the rural/village group (66.7--71.4\%) than in the two urban groups (86.5--93.9\%) on those two measures; neither survives Bonferroni correction within this family of four, and the rural subgroup's small size ($n=22$) limits how much weight either result can carry. We report this directionally rather than as a finding: the same rural/urban gap that fails to reach corrected significance here is also the sampling skew this study already discloses as a limitation, so we do not treat an uncorrected, underpowered result on the underrepresented side of that skew as evidence either way.

\begin{figure}[h]
    \centering
    \includegraphics[width=0.75\linewidth]{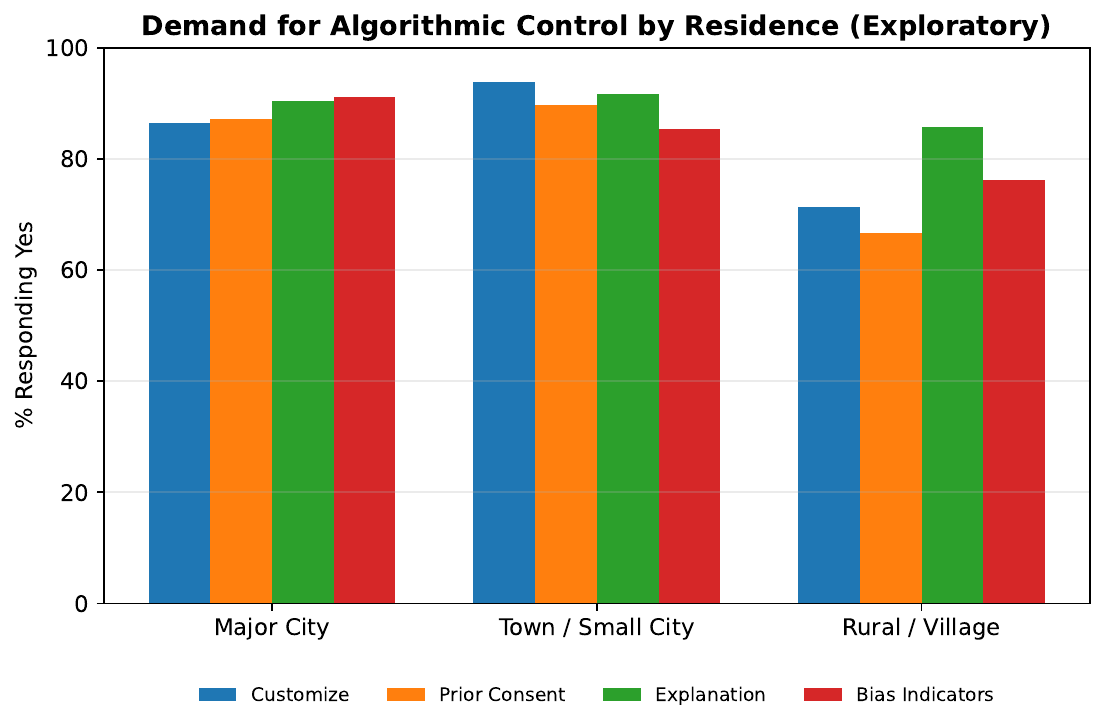}
    \caption{Support for user-control mechanisms by residence (exploratory; none survive Bonferroni correction).}
    \Description{Grouped bar chart showing support for the same four user-control mechanisms broken down by major city, town or small city, and rural/village residence. The rural/village group shows visibly lower bars on customize, prior consent, and bias indicators than the two urban groups, and a comparable bar on explanation; none of the four gaps survive Bonferroni correction.}
    \label{fig:rq4_by_residence}
\end{figure}

\subsection{Finding 4: LLM Cultural Bias Concerns}

\subsubsection{Most engagement with this question is a direct ``no,'' not a skipped explanation}

Most participants (63.5\%) already knew language models can show cultural or ideological bias. A separate, open-ended item then asked which cultures or languages get excluded from LLM responses more often, and why --- a conditional question (``if yes, why?''), not a demand that every respondent produce a reason. Of the 66 who answered anything at all (133 left the item blank entirely, itself the larger non-response fact), 34 (52\%) directly answered ``no'' (in English or Bangla) or ``no comment'' or a single punctuation mark, which is a complete, valid answer to a conditional item, not a skipped elaboration; a further roughly 10 affirmed that exclusion happens but did not say why; and 18 gave a substantive reason (one further response, a pasted LLM output, was excluded as non-authentic; \Cref{sec:methodology}).

We do not read the 34 ``no'' answers as evidence of an articulation gap --- they are a direct, complete answer to what was asked. The narrower, better-supported pattern is among the roughly 28 respondents who affirmed exclusion: about a third of that group did not go on to explain why, even though a reason was directly invited. We hold even that narrower reading loosely: an optional free-text item late in a long bilingual survey also invites non-response from fatigue or typing burden, independent of whether a respondent could explain their view if asked directly, and this survey cannot separate those explanations from each other. (See \Cref{subsec:qualitative-results} for examples.)

\subsubsection{Concerns cluster around specific themes}

Open-ended responses about cultural bias clustered around three themes: Western-centric defaults in model outputs, friction against religious or cultural values, and uncertainty about whether bias comes from training data or deliberate design. The split between data-imbalance explanations and design-choice explanations matters in practice. Fixing each requires different interventions. A minority of responses pushed back against the premise of inherent bias, or located bias in how a response is delivered (framing, tone) rather than in its content --- see \Cref{subsec:qualitative-results}.

\subsection{Qualitative Themes}
\label{subsec:qualitative-results}

Of the 75 open-ended responses collected, 25 contained sufficiently substantive content for thematic analysis. All coded responses came from the beauty-filter and LLM scenarios. The ride-sharing scenario did not include an open-ended response for the fairness-reasoning item (\Cref{app:survey-instrument}, Scenario 1), as this item was presented in a multi-select format. Therefore, no ride-sharing excerpts are reported in this section.

\subsubsection{Beauty-Filter Themes}

Responses concerning beauty filters highlighted concerns about the internalization of appearance-related harm. Participants described how exposure to idealized or altered representations may negatively influence perceptions of natural appearance. For example, one participant stated:
\begin{quote}
``They struggle to accept their natural face.''
\end{quote}

Another response explicitly identified a racial dimension of this harm:
\begin{quote}
``Black people faces problem.''
\end{quote}

These responses suggest that participants perceived beauty-filter technologies as potentially reinforcing negative self-perceptions and, in some cases, contributing to racialized standards of appearance.

\subsubsection{LLM Themes}

Responses concerning LLMs highlighted several dimensions of perceived bias, including tensions with religious and cultural values, underrepresentation in training data, and the potential influence of prompting and response framing.

Some participants expressed concerns that dominant cultural perspectives may receive greater representation than less represented cultural or religious viewpoints:
\begin{quote}
``\ldots due to western culture Muslim culture gets less importance.''
\end{quote}

Other responses attributed perceived disparities to imbalances in training data rather than deliberate favoritism by the model:
\begin{quote}
``\ldots some cultures and languages, like Bengali or indigenous ones, are often underrepresented in LLM responses due to biased training data favoring dominant languages.''
\end{quote}

However, not all participants viewed bias as an inherent or unavoidable characteristic of LLMs. One respondent emphasized the potential role of appropriate prompting in reducing biased outputs:
\begin{quote}
``\ldots if prompted correctly at the beginning, these will even give responses being non-biased towards the other missing cultures and languages.''
\end{quote}

Another response focused on how information is presented rather than solely on the content itself, suggesting that perceived bias may arise when responses lack balance:
\begin{quote}
``A response may feel biased if it presents only one perspective, uses emotionally charged language, or favors a particular group or idea without balance.''
\end{quote}

Overall, the qualitative findings indicate that participants conceptualized AI-related bias through multiple dimensions, including internalized harm, cultural and religious underrepresentation, training-data imbalance, prompting, and response framing.

\begin{table}[h]
\centering
\caption{Primary Comparisons and Correlations, with 95\% Confidence Intervals}
\label{tab:ci-summary}
\begin{tabular}{lrrr}
\toprule
\textbf{Comparison} & \textbf{Estimate} & \textbf{95\% CI} & \textbf{$n$} \\
\midrule
Ride-share: emergency $-$ casual (mean diff.) & 0.17 & [0.04, 0.30] & 185 \\
Beauty filters: self-image $-$ social pressure (mean diff.) & 0.51 & [0.28, 0.73] & 197 \\
Beauty filters: personal affect $\to$ self-image ($r$) & .545 & [.44, .64] & 197 \\
Ride-share: harm--trust correlation (Spearman $\rho$) & .599 & [.50, .68] & 199 \\
Ride-share: technical familiarity--harm (Spearman $\rho$) & .217 & [.08, .35] & 199 \\
\bottomrule
\end{tabular}
\end{table}

The two mean differences and the harm-related regression above are this study's three pre-specified primary comparisons (\Cref{sec:methodology}); the two correlations at the bottom are reported alongside them here as a compact reference for effect-size uncertainty, not as additional pre-specified primary results. Every interval excludes zero, but two of the five (the ride-share context effect and the technical-familiarity correlation) have a lower bound close to zero, consistent with the underpowered and wide-interval caveats noted for each at its point of use above.

\subsection{Summary}

In brief: context shapes fairness judgments in ride-sharing pricing; beauty-filter harm tracks with self-image more than with named social pressure; demand for transparency, consent, and control is broad and largely unmoored from demographic background; and among participants who notice LLM cultural bias specifically, many cannot say why when asked. The caveats attached to each finding --- effect size, item wording, sample composition --- are stated where the finding is reported above and are not repeated here.

The next section works through what these findings mean for system design, policy, and future research.

\section{Discussion}
\label{sec:discussion}

Fairness, in our data, is not a single fixed standard. It depends on the situation, the stakes, and who is affected. Here is what each finding means for design and policy.

\subsection{What the Context Effect Tells Us}

Participants rated emergency pricing as less fair than casual pricing, even though the price increase was identical. That suggests people are applying an ethical framework about exploitation --- not just running a calculation of cost versus benefit.

The effect is small, near this sample's own detection floor (\Cref{sec:results}), so we read it as suggestive rather than settled. It is also unevenly distributed: middle-income participants show the largest emergency-casual gap, while lower- and upper-income participants differentiate much less between the two contexts. Participants already view surge pricing critically overall; context sharpens that view rather than producing it, at least among those who differentiate between contexts at all.

A supplementary driver sample showed no significant context effect and rated the same pricing scheme fair overall more often than passengers did, though that comparison is confounded by different consent models, recruitment channels, and instrument wording between the two samples, and by the driver instrument itself asking drivers to answer from the passenger's perspective (\Cref{sec:results}). One plausible reading, which this design cannot confirm over those confounds, is that drivers and passengers sit on different sides of the same pricing system --- drivers earn from surge pricing, passengers pay it. Prior work finds personal stake can outweigh broader fairness concerns \citep{wang2020factors}, and fairness preferences track perceived risk \citep{haider2024}.

\textbf{Design implication.} Pricing systems that ignore context can feel exploitative. Systems in domains with variable stakes should detect vulnerable-user contexts and adjust. For ride-sharing, that might mean caps, or different pricing logic during emergencies.

\textbf{Policy implication.} Regulation should consider context-specific harms, not just aggregate fairness statistics. A pricing model that looks fair on average can still produce exploitative outcomes in specific situations.

\subsection{What the Beauty-Filter Finding Shows}

Beauty-filter harm tracks with self-image more than with obvious social pressure. Feeling personally affected by filters tracks closely with reduced self-confidence, on our reading of the impact scale's direction --- inferred from context rather than a labeled anchor (\Cref{app:survey-instrument}).

This is a harm that current fairness metrics simply miss. Outcome-based measures check whether different groups receive similar results. They do not capture how systems affect how people see themselves. Qualitative responses described struggling to accept one's unfiltered appearance, and, in one case, an explicitly racialized experience.

\textbf{Design implication.} Evaluation frameworks should include psychological-impact measures alongside distributional fairness. A filter can look technically fair --- similar effects across groups --- and still cause real harm.

\textbf{Policy implication.} Beauty-filter regulation should consider psychological harm, not just accuracy or demographic parity. Transparency about filter effects, and how to turn them off, matters.

\subsection{What Widespread Demand for Control Means}

Support for transparency, consent, and user control is broad (85.7--90.3\%) and flat across every subgroup we measured. Income, gender, and prior awareness of algorithmic bias all fail to predict it at conventional significance, though aware participants lean consistently, if not significantly, higher on every measure (\Cref{sec:results}).

Three further, exploratory checks (education, occupation, residence; \Cref{sec:results}) reinforce that pattern rather than complicate it, with one partial exception: rural/village participants show consistently lower raw support on three of four measures, crossing the uncorrected .05 threshold on two, though neither survives Bonferroni correction and the rural subgroup ($n=22$) is small. This is the clearest place in our data where an exception to the universal-demand story can't be ruled out, and it happens to be the one subgroup this sample is worst equipped to test. That is not evidence that rural users want less control -- just a gap we cannot close with what we collected.

User control, then, looks like a basic expectation rather than a special interest tied to any subgroup we could confidently identify. Targeted protections still matter --- the ride-sharing and beauty-filter findings show real, unevenly distributed harms. But baseline transparency, consent, and control should be defaults, not add-ons.

\textbf{Design implication.} Systems should ship with transparency and control features built in. User agency should be a baseline architectural commitment.

\textbf{Policy implication.} AI governance frameworks should treat user sovereignty --- meaningful consent, explanation, and control --- as a baseline requirement, not a tiered feature set.

\subsection{What the LLM Open-Ended Responses Suggest}

Most participants recognized algorithmic bias in closed-form items. When asked directly which cultures or languages get excluded from LLM responses and why, most who answered at all said, plainly, that they did not think this happens --- a substantive position, not a failure to engage (\Cref{sec:results}). The narrower pattern worth taking seriously sits inside the smaller group who affirmed exclusion: roughly a third of them did not go on to explain why, even when a reason was directly invited. That narrower pattern looks like a gap between affirming a problem and being able to articulate it, though we hold the reading loosely: this survey cannot fully rule out ordinary non-response --- fatigue or typing burden late in a long bilingual instrument --- as a competing explanation for the same pattern.

Digital-literacy programs that focus only on technical understanding may be missing this narrower gap, where it exists. Users who recognize a specific instance of bias may still benefit from practice articulating why it matters, and connecting it to their own contexts.

\textbf{Design implication.} Platforms should help users understand not just that bias exists, but why it matters in specific domains. Educational prompts or context-specific explanations might bridge the gap.

\textbf{Education implication.} AI literacy programs should include practice analyzing concrete examples of bias and discussing why they matter --- not just abstract descriptions of how algorithms work.

\subsection{Limitations}

The sample is urban, young, male, and middle-income by Bangladeshi standards, limiting generalization to the broader population; rural and older perspectives are underrepresented, with only 22 rural/village respondents reached, the same subgroup where the exploratory residence check (\Cref{sec:results}) finds the one uncorrected, underpowered hint that this study's flat, universal-demand finding might not hold.

Scenario-based methodology enables controlled comparison but cannot capture the messiness of real algorithmic encounters, and self-reported perception may miss unconscious bias acceptance or adaptation that only longitudinal or behavioral designs would catch.

The self-image/social-pressure item pair's low Spearman-Brown-corrected reliability ($\alpha=.427$, $r\approx.27$) is not a defect: \Cref{sec:results}'s finding is specifically that these two deliberately contrasted items diverge, so a low correlation is the expected result, not evidence of a scale that fails to measure what it intends.

Several further caveats apply. No minor-specific consent or assent procedure was applied to the fourteen under-18 respondents, including on the beauty-filter scenario's self-image items (\Cref{sec:methodology}). The RQ4-by-income chi-square tests, and the exploratory by-education, by-occupation, and by-residence families added later, all have expected cell counts below the conventional adequacy threshold, given how few upper-income and other minority-subgroup respondents this convenience sample reached (\Cref{sec:results}); we report them for completeness but weight them less than the by-gender and by-awareness results, which do meet the threshold. The beauty-filter impact items' response-scale direction is inferred from context rather than labeled anchors (\Cref{app:survey-instrument}). And three design confounds, each detailed at its point of use in \Cref{sec:results}, remain unresolved: the emergency and casual pricing items were always shown in the same fixed order; the self-image and social-pressure items differ in grammatical person as well as construct, so a third-person effect is a live alternative explanation for Finding 2; and all four RQ4 control-demand items are unipolar and socially desirable with no reverse-worded item, so Finding 3's near-universal support cannot be fully distinguished from acquiescence bias.

\subsection{Implications for Global South Contexts}

This study adds a South Asian perspective to fairness research that has mostly been built from Western findings, and three considerations here may travel beyond Bangladesh. One is contextual moral reasoning --- people judge fairness against the situation, not just the outcome, and exploitation and vulnerability carry real weight in that judgment. A second is internalized psychological harm: some harms sit below conscious recognition entirely, in a place distributional fairness metrics were never built to look. The third is closer to a baseline expectation than a preference --- support for transparency and user agency that holds regardless of who is asked.

Regulatory approaches built around user-agency rights track how users actually think about fairness better than paternalistic protection models do. Governance frameworks built mainly from Western contexts risk reproducing the same coverage gaps.

Set against the closest prior South Asian work (\Cref{sec:literature-review}), what looks distinctive here is narrower than a general ``Global South perspective.'' \citet{ramesh2022} find that Indian instant-loan borrowers' ability to contest algorithmic decisions is shaped by structural power relations with the platform, not primarily by information or awareness gaps; our RQ4 finding is compatible with that --- support for control mechanisms here is close to universal and does not track technical awareness either --- but we did not test power relations directly, so we can describe the same pattern, not extend their causal account. \citet{santos2020} study Uber price-fairness perceptions directly and give this paper's ride-sharing finding a closer predecessor than the general fairness-perception literature; what this study adds on top of that base is the emergency-versus-casual context manipulation itself, which \citet{santos2020} does not report, plus the additional beauty-filter and LLM domains outside pricing altogether.

Future work should test whether these patterns hold in other Global South contexts, and how they translate into concrete design and policy interventions.

\subsection{A Reproducible Methodology}

This study contributes more than findings. We wanted to offer researchers doing similar scenario-based fairness-perception work --- especially under-resourced teams, or teams in under-studied regions --- a practical template, so we release the complete pipeline: de-identified data, a column-level codebook, a single re-runnable analysis script that regenerates every statistic and figure here, and documented inclusion criteria.
\section{Future Work}
\label{sec:future-work}

This study gives an initial evidence base for fairness perceptions in Bangladesh. Data collection is closed for the project. The remaining directions below all require new data collection or are independent of the survey data entirely.

\subsection{A Longitudinal Extension}
\label{subsec:longitudinal}

This study's design is cross-sectional. Every measure is a single snapshot, which leaves several of its own findings unable to distinguish a stable trait from a process that unfolds, compounds, or fades over time. Four specific longitudinal questions follow directly from the results in \Cref{sec:results} --- not from longitudinal research in the abstract:

\begin{itemize}
    \item \textbf{Does beauty-filter self-image harm actually compound with exposure?} This survey instrument has no item measuring filter-use frequency at all (\Cref{sec:results}), so the cross-sectional relationship reported here ($R^2=.30$ between two self-reported impact measures) cannot speak to dose-response. It cannot separate a genuine erosion effect from a stable individual difference that predicts both filter use and self-image sensitivity. Answering this needs a design that measures actual exposure for the first time --- not just repeated measurement of the impact items already in this instrument.
    \item \textbf{Does the ride-sharing exploitation framework strengthen, habituate, or fade with repeated exposure to surge pricing?} This study's small, single-exposure context effect (\Cref{sec:results}), suggestive rather than settled at this sample size, is a snapshot regardless. Repeated real-world exposure could plausibly sharpen moral judgment (cumulative outrage) or blunt it (normalization). Those predict opposite policy responses.
    \item \textbf{Is the demand for algorithmic transparency and control (\Cref{sec:results}) a stable baseline preference, or does it track salient events} --- a publicized AI incident, a new LLM release --- rising and falling with an issue-attention cycle rather than remaining constant?
    \item \textbf{Does actually receiving the transparency and control mechanisms respondents asked for change downstream trust and behavior}, or does stated preference for these mechanisms fail to predict revealed behavior once they are available? That gap is well documented elsewhere in privacy and consent research.
\end{itemize}

A practical design combining a repeated-measures panel (the core scenario battery re-administered at intervals, testing question 3 and providing repeated-measures data for questions 1--2), a shorter-cycle diary or experience-sampling component for beauty-filter use specifically (testing question 1's causal direction with within-person, event-level data rather than retrospective self-report), and a small quasi-experimental arm giving a subset of participants a working transparency/control prototype against a control group (testing question 4, and building on the prototyping direction below) would address all four questions with one coordinated data-collection effort. A fully worked protocol --- proposed wave timing, retention strategy, instrumentation, and analysis approach (cross-lagged panel and growth-curve models are the natural fit for questions 1--3) --- is maintained as a companion planning document alongside this manuscript. It is kept out of the main text so it can be revised independently as a concrete study is scoped.

\subsection{Other Directions Requiring New Data Collection}
\begin{itemize}
    \item \textbf{Expanding demographic and contextual coverage.} Extending the sample to rural populations, older adults, and lower digital-literacy groups would address the sampling skew noted in \Cref{sec:discussion}, using a formal probability or stratified design rather than convenience recruitment.
    \item \textbf{Stakeholder-perspective research: drivers as well as passengers.} A small, driver-recruited supplementary sample (\Cref{subsec:followup-results}) hints that ride-sharing fairness may look different from the supply side of the same pricing algorithm --- drivers in that sample were significantly more likely to call differential pricing fair overall than the primary, passenger-oriented sample. The sample is too small and non-representative to draw conclusions about drivers as a group. It does point to a distinct, promising direction: a study explicitly comparing driver and passenger fairness perceptions, ideally testing self-interest and risk perception directly as mediating variables rather than inferring them after the fact, building on \citet{wang2020factors} and \citet{haider2024}.
    \item \textbf{Cross-cultural replication.} Repeating this study's scenario battery in other Global South contexts (South and Southeast Asia, Sub-Saharan Africa, Latin America) would test whether the findings here --- the exploitation framework, the demographically-unmoderated transparency demand, the internalized-harm mechanism --- are Bangladesh-specific or reflect broader patterns across similarly-positioned digital economies. This extends the study's own core argument: Global South perspectives are not interchangeable with each other any more than they are with Western ones.
    \item \textbf{Experimental designs isolating causal mechanisms.} This study's contextual and demographic effects are observational. Between-subjects experiments that manipulate a single factor (randomly assigning identical pricing scenarios framed as emergency vs.\ routine) would isolate causal effect sizes more precisely than the within-subjects comparisons reported here.
    \item \textbf{Additional algorithmic domains.} Extending beyond ride-sharing, beauty filters, and LLMs to higher-stakes domains increasingly relevant to Bangladesh's digital governance trajectory --- credit scoring, algorithmic hiring, e-government service allocation --- would test whether the moral-exploitation framework and universal-transparency findings generalize to domains with larger material stakes.
    \item \textbf{Labeled anchors for the beauty-filter impact items.} The self-image, social-pressure, and personal-affect items (\Cref{app:survey-instrument}) carry no labeled scale endpoints in the live instrument, so this study's own interpretation of ``higher = more harm'' is inferred from context rather than guaranteed (\Cref{sec:discussion}). Unlike the other items in this subsection, this direction needs new data collection specifically: a short follow-up survey re-asking the same items with explicit anchors, on a fresh sample, would let a future analysis confirm or correct that interpretation directly. We do not recommend combining the self-image and social-pressure items into a single composite in the meantime: \Cref{sec:results}'s finding is specifically that they diverge, and their low Spearman-Brown-corrected reliability coefficient ($\alpha=.427$, implying an inter-item $r\approx.27$) is evidence for that divergence, not a defect to average away.
\end{itemize}

\subsection{Independent of Survey Data: Technical and Design Work}
\begin{itemize}
    \item \textbf{Independent technical bias audits.} Direct algorithmic audits of beauty-filter tools popular in Bangladesh (measuring skin-tone shift or feature modification directly from filter output, following the audit methodology of \citet{riccio2023}) would complement this study's self-report findings with independent technical measurement. They do not require additional survey responses.
    \item \textbf{Cross-platform governance comparison.} Comparing publicly available fairness and transparency features across ride-sharing platforms operating in Bangladesh (Uber, Pathao, Obhai) could be conducted from public documentation. No new survey data needed.
    \item \textbf{Prototyping fairness tools.} Interfaces such as fairness dashboards or browser extensions that surface the transparency and control mechanisms respondents asked for (\Cref{sec:results}) are a natural next design step. They would directly supply the intervention arm described in \Cref{subsec:longitudinal}. A companion practitioner-facing document translates this study's headline findings into concrete interaction-design guidance for teams building on this work.
    \item \textbf{Participatory co-design and policy translation.} Structured workshops with users, developers, and policymakers to translate the findings in \Cref{sec:discussion} into concrete transparency and control guidelines, and into evidence-based recommendations for Bangladeshi digital policy and AI-ethics guidance.
\end{itemize}

Pursued in this order --- completing analysis of the current dataset first, then independent technical/design work and cross-sectional replication, with the longitudinal extension and other new-collection directions as the longer-term agenda --- this roadmap moves from the empirical findings reported here toward practical design, policy application, and a genuinely causal and time-resolved understanding of algorithmic fairness perceptions. It does that without requiring resources this project does not currently have.

\section{Conclusion}
\label{sec:conclusion}

This study examined how 199 Bangladeshi participants judge fairness across three algorithmic domains: ride-sharing pricing, AI beauty filters, and large language models. Four patterns emerged.

Context shapes fairness judgments. The same algorithmic behavior gets judged differently depending on the situation --- emergency pricing feels less fair than casual pricing, even when the increase is identical.

A related but distinct pattern shows up with beauty filters: harm operates mainly through internalized effects on self-image, not through obvious social pressure, and outcome-based fairness metrics do not capture this at all.

Demand for transparency, consent, and user control, meanwhile, is widespread and cuts across the demographic lines we could measure (\Cref{sec:results}). These should be defaults, not optional features.

Finally, among participants who report specific instances of LLM cultural bias, a meaningful minority struggle to explain why it matters when invited to.

\subsection{What This Means for Practice}

System designers get three concrete directions out of this: context-aware logic that detects vulnerable-user situations and adjusts accordingly, transparency and control features built in as defaults, and evaluation frameworks that weigh psychological impact alongside distributional fairness. Policymakers get a narrower but sharper one --- AI governance should treat user sovereignty (meaningful consent, explanation, control) as a baseline requirement, and regulations should weigh context-specific harms rather than aggregate statistics alone. And for researchers, this adds a Global South perspective to fairness research, plus a reproducible methodology other teams can adapt directly.

\section*{Data and Code Availability}
\label{sec:data-availability}
\addcontentsline{toc}{section}{Data and Code Availability}

Code, data, and survey instrument materials are publicly available at \url{https://github.com/ahmed-shafi/fairness-perceptions-global-south}. The repository includes the quantitative analysis pipeline (Python, pandas, scipy.stats, statsmodels), a column-level codebook mapping every survey item to the variable names used in this paper, de-identified response data for both the primary sample and the supplementary follow-up sample, and the full survey instrument in English and Bangla, including the informed-consent text and scenario vignettes. The follow-up sample's identifying fields (name, and optionally email and signature; \Cref{subsec:followup-sample}) are redacted in place before release and never published in any form. Running the pipeline against the released data regenerates every quantitative statistic, table, and figure reported in \Cref{sec:results} from scratch.

\section*{Author Contributions}
Ahmed Abdal Shafi Rasel: Conceptualization, Methodology, Writing -- original draft, Writing -- review \& editing, Supervision. Ahmed Mustafa Amlan: Data curation, Formal analysis, Writing -- original draft. Tasmim Shajahan Mim: Data curation, Formal analysis, Writing -- original draft.

\section*{Acknowledgments}
Generative AI was used solely to assist with English language editing, including grammar, spelling, and clarity, as well as LaTeX formatting and document structuring. Claude (Anthropic) was used to assist with these tasks. Generative AI was not used for study design, data collection, statistical analysis, or interpretation of the results. The authors reviewed and edited all AI-assisted content and take full responsibility for the content of the publication.

\section*{Ethics and Privacy Statement}

The primary survey (\Cref{sec:methodology}) collected no personally identifying information: no names, phone numbers, email addresses, or IP addresses. All participants gave informed consent. Fourteen respondents were under 18; no minor-specific consent or assent procedure beyond the standard informed consent process was applied to this subgroup. The supplementary driver sample (\Cref{subsec:followup-sample}) used a separate consent model that did collect a name and, optionally, an email address and signature; unlike the primary survey, this sample is not free of identifying information, though those fields are dropped before analysis and never appear in any released output (Data and Code Availability section).

\bibliographystyle{plainnat}
\bibliography{reference}

\appendix
\refstepcounter{section}
\setcounter{subsection}{0}
\phantomsection
\label{app:survey-instrument}

\clearpage
\begin{center}
{\Huge\bfseries Appendix A: Full Survey Instrument\par}
\end{center}
\bigskip

\noindent\textit{Bangla wording supplement.} Appendix~B, reproduced at the end of this arXiv version, records the Bangla item wording, consent text, and scenario vignettes for the primary survey. The English instrument below remains the canonical presentation in this manuscript. Item-level wording is recoverable from the primary survey export (\texttt{survey.csv}); the consent text and scenario vignettes are not preserved in that export and are instead transcribed directly from the live survey instrument.

\subsection{Informed Consent}
You are invited to take part in a research survey about how people perceive fairness in digital decision-making systems. Your participation is completely voluntary, and you may skip any question or withdraw at any time. We will not collect any personally identifying information, and all your responses will remain confidential. The results will only be used for academic research and analysis.

\begin{enumerate}
    \item \textbf{Do you agree to participate in this research survey about fairness in digital decision-making systems?} Yes / No
\end{enumerate}

\subsection{Demographic Questions}

\begin{enumerate}
    \item \textbf{Age:} Under 18 / 18--20 / 21--23 / 24--26 / 27--30 / Above 30
    \item \textbf{Gender:} Male / Female
    \item \textbf{Education Level:} SSC or below / HSC / Bachelor's / Master's
    \item \textbf{Occupation:} Student / Job Holder / Self-employed / Business / Unemployed
    \item \textbf{Medium of Study:} What was the primary medium of instruction in your education? Bangla Medium / English Medium / English Version / Madrasa Education
    \item \textbf{Field of Study or Work:} Computer Science / Engineering; Business / Economics; Social Sciences / Humanities; Natural Sciences; Health / Medical; English
    \item \textbf{Hometown Background:} Major City / Town or Small City / Rural Area / Village
    \item \textbf{Socioeconomic Background:} Lower-income / Middle-income / Upper-income
    \item \textbf{Technology Familiarity:} Very Low / Low / Moderate / High / Very High
    \item \textbf{Algorithmic Awareness:} Have you heard of ``algorithmic bias'' or ``AI fairness'' before? Yes / No
    \item \textbf{What does ``Algorithm Fairness'' mean to you?} Equal treatment for everyone; avoiding bias against any group; transparency and explainability of algorithms; regular testing and updating to ensure fairness; prioritizing accuracy over fairness
    \item \textbf{Language Proficiency:} Bangla / English
\end{enumerate}

\subsection{Scenario 1: Ride-Sharing Fare Bias}
Suppose you and a friend both open a ride-sharing app at the same time to book a ride from the same pickup point to the same destination. Surprisingly, you see a fare of 320 BDT while your friend sees 250 BDT. Both of you have almost identical ride histories. Later, you find out that the app uses personal data -- such as your phone model, location history, and previous spending behavior -- to personalize prices.

\begin{enumerate}
    \item Were you aware ride-sharing apps might show different prices to different users? (Yes/No)
    \item Do you think this pricing is fair overall? (Fair/Unfair)
    \item Why do you think this is unfair? (select all that apply: discriminates against low-income areas; exploits urgency or emergencies; increases inequality; reduces trust; I think this is fair)
    \item How much do you feel harmed personally by this price difference? (1--5)
    \item If this is common, how would it affect your trust in the platform? (1--5)
    \item How do you feel when fare increases during rain or stormy weather? (1--5)
    \item During holidays like Eid when demand is high, do you think the price hike is acceptable? (Yes/No)
    \item If someone from a low-income area pays less than someone from a high-income area for the same ride, how fair do you find this? (1--5)
    \item Imagine you urgently need a ride to the hospital during a medical emergency and see the price has increased by 20\%. How fair do you find this? (1--5)
    \item Imagine you are booking a ride for a casual visit and see the price has increased by 20\%. How fair do you find this? (1--5)
    \item Is personalized pricing more acceptable in e-commerce than in ride-sharing? (Yes/No)
    \item Would offering discounts to people with low incomes or in critical situations make this feel fairer? (Yes/No)
    \item Do you think ride-sharing apps should explain how your fare is calculated? (Yes/No)
    \item If apps are transparent about how pricing works, would that make this option feel fairer? (Yes/No)
\end{enumerate}

\subsection{Scenario 2: Beauty Filter Bias}
Suppose you're experimenting with several beauty filters on Instagram or Snapchat for fun. You notice that most filters automatically lighten your skin tone, smooth your face, and change your features to match a particular beauty ideal. Later, a friend from a different ethnic background tries the same filters, and the results look dramatically different on them.

\begin{enumerate}
    \item Were you aware that beauty filters often lighten skin, smooth features, and promote specific beauty standards? (Yes/No)
    \item Do you know what the current beauty standards promoted on social media are? (Yes/No)
    \item Do you think these filters are fair across different skin tones and facial features? (Yes/No)
    \item If unfair, why? (select all that apply: promotes unrealistic beauty ideals; discriminates against darker skin tones; causes social pressure; creates economic pressure to buy beauty products; no, I think it's fair)
    \item How personally affected would you feel by such filters? (1--5, unlabeled endpoints)
    \item How do such filters affect confidence or self-image? (1--5, unlabeled endpoints)
    \item Are some groups more negatively affected? (Yes/No; if yes, open-ended follow-up)
    \item People often react more positively to filtered photos. Do you feel social pressure to use beauty filters to fit in? (1--5, unlabeled endpoints)
    \item If someone is using filters professionally (e.g., for job applications, modeling), do you think fairness standards should be stricter? (Yes/No)
    \item Is this more problematic in beauty filters or fun filters? (Beauty filters / Fun filters / Both equally)
    \item Would transparency labels make filters feel fairer? (Yes/No)
    \item Should filters work equally well across all ethnicities? (Yes/No)
\end{enumerate}

\emph{A note on the three 1--5 items above:} unlike the ride-sharing fairness items in \Cref{app:survey-instrument}, Scenario 1 (interpreted directionally in \Cref{sec:methodology}'s codebook as 1=least fair to 5=most fair), the three beauty-filter impact items above carry no labeled endpoints in the live instrument. \Cref{sec:results} and \Cref{sec:discussion} interpret a higher score as greater self-reported impact/harm, consistent with this scenario's framing (filters described as lightening skin tone and reshaping features toward one ideal) and with the direction of the correlation between the personal-affect and self-image items; but that interpretation is inferred from context rather than guaranteed by the scale itself, and we flag it as a limitation of the instrument rather than present it as unambiguous.

\subsection{Scenario 3: Language Model Cultural Bias}
Imagine you ask a language model a question about family values. The answer you receive focuses mainly on individualism, personal freedom, and independence. However, it doesn't mention core cultural values like respect for elders or religious traditions, which are important in your community. Later, you find that your friend received a very different response that aligns more closely with their cultural background.

\begin{enumerate}
    \item Did you know that large language models (LLMs) can reflect cultural or ideological biases? (Yes/No)
    \item How fair do you think it is if the values of one culture dominate the responses of a language model? (1--5)
    \item What aspects of the LLM's response made it feel biased or unbiased to you? (select all that apply: agreed too quickly without justification; avoided challenging an incorrect belief; favored a particular culture, group, or opinion; treated all sides fairly and equally) -- documented as fixed-choice only, but 5 of 192 respondents in the live data answered in free text instead (\Cref{sec:methodology})
    \item Have you ever felt misunderstood because the LLM failed to grasp your cultural background or context? (Yes/No)
    \item Would repeated bias in LLM responses reduce your trust in the technology? (1--5)
    \item To what extent do you believe biased answers from LLMs impact cultural identity? (1--5)
    \item Do you think certain cultures or languages are excluded more often in LLM responses? If yes, why? (open-ended)
    \item Have you ever felt that a response from an LLM overlooked or disregarded important cultural values? (Yes, I have experienced this / No, I have not experienced this)
    \item In your view, should cultural fairness hold greater importance in value-based answers compared to factual ones? (Yes/No)
    \item Do you think allowing users to select their cultural context would improve fairness in LLM responses? (1--5, Strongly agree--Strongly disagree)
    \item If an AI assistant changes its answer to align with your personal opinion, do you think that reflects cultural bias? (Yes/No)
    \item Should AI always agree with the user's cultural viewpoint? (Yes, to show respect / No, it should remain neutral / Depends on the context)
\end{enumerate}

\subsection{RQ4: User Expectations (All Scenarios)}

\begin{enumerate}
    \item What things should the system show or let you do so you think is fair? (select all that apply: show how prices/results are decided; let me change or customize settings; allow me to report unfair behavior or results; provide clear rules or policies; show data or history related to me)
    \item Would you like to customize how the system works for you? (Yes/No)
    \item Should platforms explain how their system works? (Yes/No)
    \item Should the system ask for your preferences before making decisions that affect you? (Yes/No)
    \item Should AI platforms clearly indicate when an answer may be biased or incomplete for your culture? (Yes/No)
\end{enumerate}

\subsection{Supplementary Follow-Up Survey (Ride-Sharing Only)}
\label{app:followup-instrument}

This instrument was fielded separately (24 Aug -- 10 Dec 2025, a window entirely inside the primary survey's own collection period rather than after it closed; \Cref{subsec:followup-sample}), recruited primarily from ride-sharing drivers and covering ride-sharing only. Unlike the primary instrument above, this form also requested a full name, an email address, and a signature; those fields are omitted here since they fall outside the substantive scenario and were dropped from all analysis before it began (\Cref{subsec:followup-sample}). Two items add concrete examples not present in the original wording, marked as such below; the rest are identical to \Cref{app:survey-instrument}'s Scenario 1.

\subsubsection{Demographic Questions}
\begin{enumerate}
    \item \textbf{Age:} 21--23 / 24--26 / 27--30 / Above 30
    \item \textbf{Gender:} Male / Female
    \item \textbf{Hometown Background:} Major City / Town or Small City / Rural Area / Village
    \item \textbf{Algorithmic Awareness:} Have you heard of ``algorithmic bias'' or ``AI fairness'' before? Yes / No
\end{enumerate}

\subsubsection{Scenario: Ride-Sharing Fare Bias}
Same framing scenario as \Cref{app:survey-instrument}, Scenario 1.

\begin{enumerate}
    \item Were you aware ride-sharing apps might show different prices to different users? (Yes/No)
    \item Do you think this pricing is fair overall? (Fair/Unfair)
    \item Why do you think this is unfair? (select all that apply: discriminates against low-income areas; exploits urgency or emergencies; increases inequality; reduces trust; I think this is fair)
    \item How much do you feel harmed personally by this price difference? (1--5)
    \item If this is common, how would it affect your trust in the platform? (1--5)
    \item How do you feel when fare increases during rain or stormy weather? (1--5)
    \item During holidays like Eid when demand is high, do you think the price hike is acceptable? (Yes/No)
    \item \emph{[Wording adds concrete examples.]} If someone from a low-income area (e.g., Kerani Ganj) pays less than someone from a high-income area (e.g., Gulshan) for the same ride, how fair do you find this? (1--5)
    \item Imagine you urgently need a ride to the hospital during a medical emergency and see the price has increased by 20\%. How fair do you find this? (1--5)
    \item Imagine you are booking a ride for a casual visit and see the price has increased by 20\%. How fair do you find this? (1--5)
    \item Is personalized pricing more acceptable in e-commerce than in ride-sharing? (Yes/No)
    \item \emph{[Wording adds concrete examples.]} Would offering discounts to people with low incomes or in critical situations (such as emergencies or financial hardship) make this feel fairer? (Yes/No)
    \item Do you think ride-sharing apps should explain how your fare is calculated? (Yes/No)
    \item If apps are transparent about how pricing works, would that make this option feel fairer? (Yes/No)
\end{enumerate}

\includepdf[pages=-,pagecommand={\thispagestyle{plain}}]{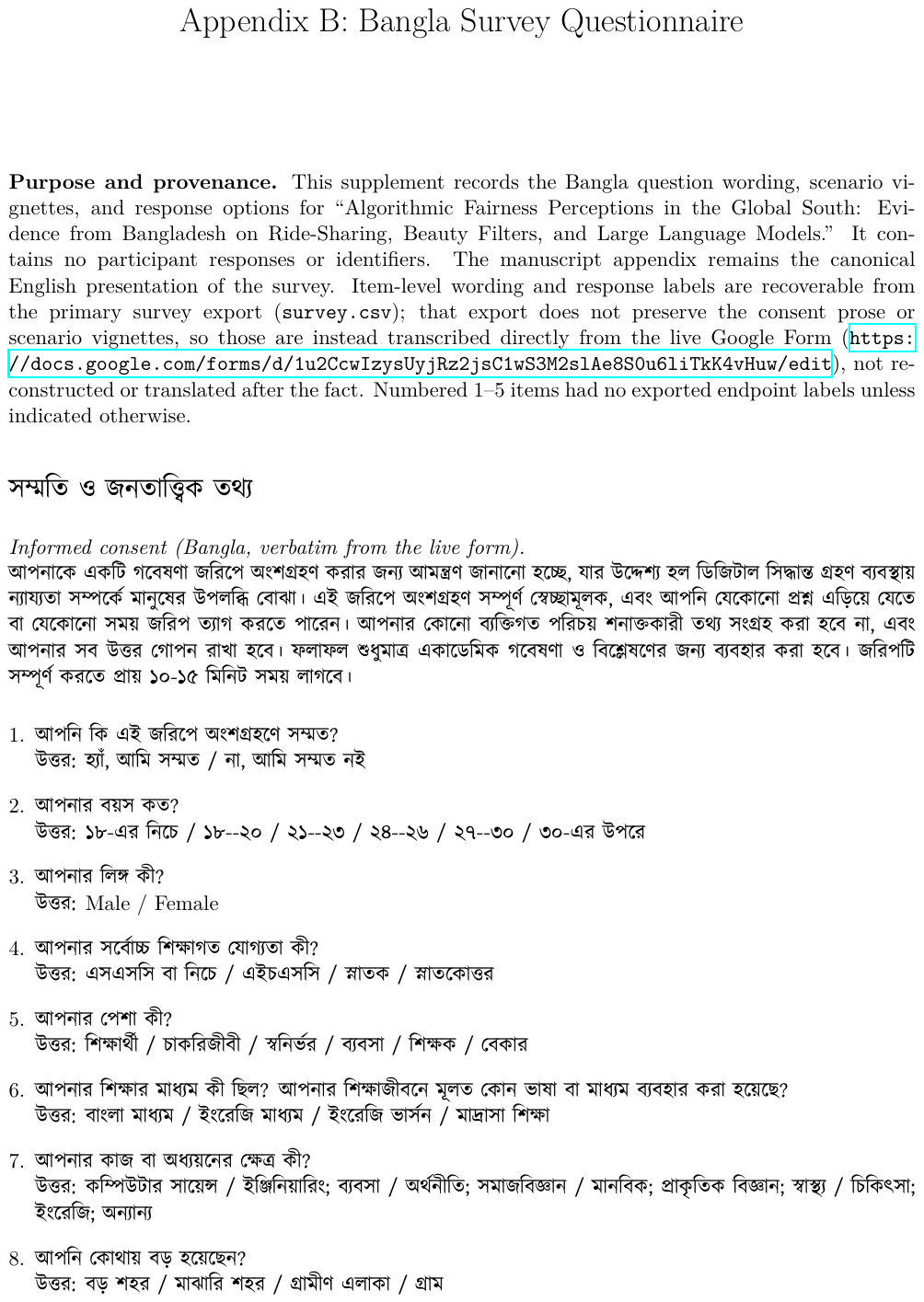}

\end{document}